\documentclass{PoS}

\title{Baryon interactions from lattice QCD with physical masses
  ---$S=-3$ sector: $\Xi\Sigma$ and $\Xi\Lambda-\Xi\Sigma$---}

\ShortTitle{Baryon interactions from lattice QCD with physical masses
  ($S=-3$ sector)}

\author{\speaker{Noriyoshi Ishii},$^{ab}$,
  Sinya Aoki,$^{bcd}$
  Takumi Doi,$^{b}$
  Shinya Gongyo,$^{be}$
  Tetsuo Hatsuda,$^{bf}$
  Yoichi Ikeda,$^{ab}$
  Takashi Inoue,$^{bg}$
  Takumi Iritani,$^{b}$
  Takaya Miyamoto,$^{bc}$
  Keiko Murano,$^{ab}$
  Hidekatsu Nemura,$^{bd}$ and
  Kenji Sasaki$^{bc}$
  \\
        $^a$Research Center for Nuclear Physics, Osaka university, Osaka 567-0047, Japan\\
        $^b$Theoretical Research Division, Nishina Center, RIKEN, Wako 351-0198, Japan\\
        $^c$Yukawa Institute for Theoretical Physics, Kyoto University, Kyoto 606-8502, Japan\\
        $^d$Center for Computational Sciences, University of Tsukuba, Ibaraki 305-8571, Japan\\
        $^e$CNRS, Laboratoire de Math\'ematiques et Physique Th\'eorique, Universit\'e de Tours, 37200 France\\
        $^f$iTHEMS Program and iTHES Research Group, RIKEN, Wako 351-0198, Japan\\
        $^g$Nihon University, College of Bioresource Sciences, Kanagawa 252-0880, Japan\\
        E-mail: \email{ishiin@rcnp.osaka-u.ac.jp}}

\abstract{We present  lattice QCD results of  baryon-baryon potentials
  in  $S=-3$  sector,  i.e.,   $\Xi\Sigma$  ($I=3/2$)  potentials  and
  $\Xi\Lambda$-$\Xi\Sigma$  coupled  channel potentials  ($I=1/2$)  by
  using the 2+1  flavor gauge configurations with  almost the physical
  quark masses  generated on $96^4$  lattice with $a^{-1}  \simeq 2.3$
  GeV and $L =  96a \simeq 8.1$ fm where $m_{\pi}  \simeq 146$ MeV and
  $m_K\simeq 525$ MeV.
  These potentials  are obtained based  on the time-dependent  HAL QCD
  method with a non-relativistic approximation.
  Qualitative behaviors of the results are found to be consistent with
  those in the flavor SU(3) limit.  }

\FullConference{34th annual International Symposium on Lattice Field Theory\\
		24-30 July 2016\\
		University of Southampton, UK}

\newcommand{\Fig}[1]{Fig.~\ref{#1}}
\newcommand{\Eq }[1]{Eq.~(\ref{#1})}
\newlength{\Figwidth}
\newcommand{\agt}{\mbox{\,\raisebox{0.5ex}{$>$}\hspace*{-0.7em}\raisebox{-0.5ex}{$\sim$}\,}}

\begin{document}
\section{Introduction}
One of the most important missions of J-PARC in the nuclear physics is
experimental determination of hyperon-nucleon (YN) and hyperon-hyperon
(YY) potentials. These  potentials play an important  role in studying
the  structures of  the  hyper-nuclei and  properties  of the  neutron
stars. Because of the short life time  of the hyperons, it is not easy
to determine  the hyperon potentials  for large $|S|$ sectors  even by
J-PARC, so that they focus mainly on $S=-1$ and $-2$ sectors.
On the other hand, lattice QCD determination of hyperon potentials was
recently proposed by HAL QCD collaboration, where Nambu-Bethe-Salpeter
(NBS) wave functions  are used to define the  hyperon potentials which
are        faithful        to        the        scattering        data
\cite{Ishii:2006ec,Aoki:2009ji,Aoki:2011gt,Aoki:2012tk}.
Since the  statistical noise reduces  as the number of  strange quarks
increases, lattice QCD determination becomes easier for larger $|S|$.

Here, we  present our results  of hyperon-hyperon (YY)  potentials for
$S=-3$ sectors.   We use 2+1  flavor lattice QCD  gauge configurations
generated by employing  almost the physical pion  mass $m_{\pi} \simeq
146$ MeV by using K computer at AICS \cite{Ishikawa:2015rho}.

\section{Lattice QCD setup}

We  use the  2+1 flavor  gauge configurations  at almost  the physical
point generated  by K computer at  AICS \cite{Ishikawa:2015rho}.  They
are generated on $96^4$ lattice by employing the RG improved (Iwasaki)
gauge   action  at   $\beta   =  1.82$   with  the   nonperturbatively
$O(a)$-improved   Wilson  quark   (clover)  action   at  $(\kappa_{\rm
  ud},\kappa_{\rm s})=(0.126117,  0.124790)$ with $c_{\rm SW}  = 1.11$
and the 6-APE stout smeared links  with the smearing parameter $\rho =
0.1$.  It  leads  to  the  lattice spacing  $a^{-1}  \simeq  2.3$  GeV
($a\simeq 0.085$  fm), the  spatial extension  $L=96a \simeq  8.1$ fm,
$m_{\pi} \simeq 146$ MeV, and $m_{K}\simeq 525$ MeV.
In  our  calculation,  200  gauge  configurations  are  used  for  the
measurements.
Quark  propagators are  generated  by imposing  the periodic  boundary
condition  along the  spatial direction  while the  Dirichlet boundary
condition  is imposed  on the  temporal  direction on  the time  slice
$t=t_1$ which is separated from the wall source as $t_1 - t_0 = 48$.
48 source points are used.
Forward and  backward propagations  are combined  by using  the charge
conjugation  and time  reversal symmetries  to double  the statistical
data of two-point and four-point hyperon correlators.
Each  gauge configuration  is used  4  times by  using the  hypercubic
SO(4,$\mathbb{Z}$) symmetry of $96^4$ lattice.
Statistical data  are averaged  in the  bin of the  size 10,  which is
equivalent to 100 HMC trajectories.  Jackknife prescription is used to
estimate the statistical errors.

\section{Single baryon sector}

\begin{figure}[h]
  \begin{center}
    \includegraphics[angle=-90,width=\Figwidth]{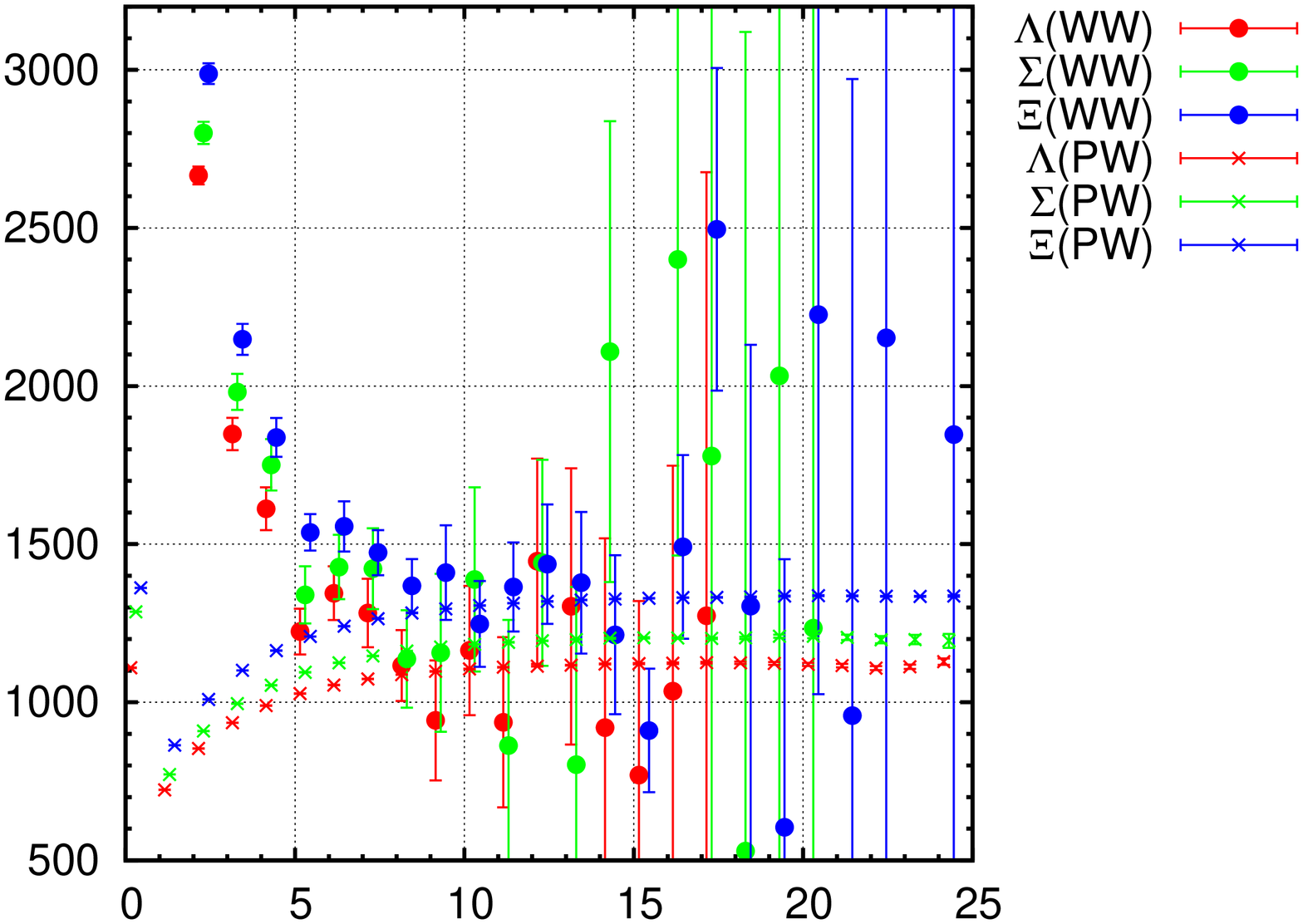}
    \includegraphics[angle=-90,width=\Figwidth]{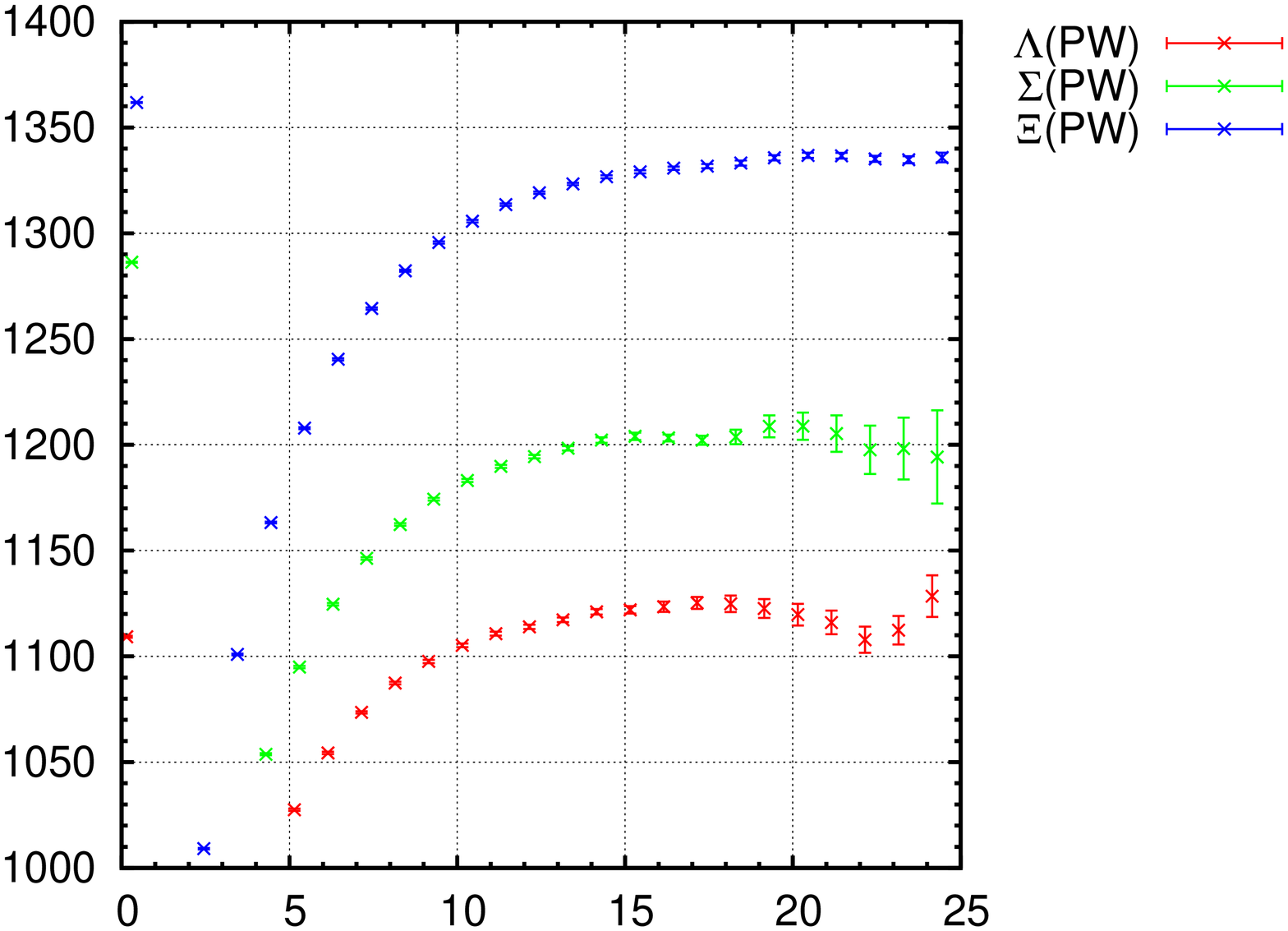}
  \end{center}
  \caption{The  effective  mass  plots  of  two-point  correlators  of
    $\Lambda$,  $\Sigma$  and  $\Xi$   for  (WW)  wall-wall  and  (PW)
    point-wall (lhs), and those for point-wall correlators (rhs).}
  \label{fig:effmass}
\end{figure}

Effective mass  plots for $\Lambda$,  $\Sigma$ and $\Xi$ are  shown in
\Fig{fig:effmass}  for the  two-point correlators  with wall-sink  and
wall-source(wall-wall)  together with  the  ones  with point-sink  and
wall-source(point-wall).
%
%
We perform  single exponential  fits simultaneously for  the wall-wall
and point-wall correlators with
\begin{equation}
  C_{\rm WW}(t) \simeq a_{\rm WW} \exp\left(-mt\right),
  \hspace{2ex}
  C_{\rm PW}(t) \simeq a_{\rm PW} \exp\left(-mt\right),
\end{equation}
by using three  fit parameters, i.e., a baryon mass  $m$, two overlaps
parameters $a_{\rm WW}$ for the  wall-wall correlator and $a_{\rm PW}$
for the point-wall correlators.
The overlap  parameters are used to  determine the Z factors  of local
composite  hyperon  operators  appearing  in the  limit  $\psi(x)  \to
Z^{1/2}\psi_{\rm  out}(x)$ as  $x_0\to  +\infty$  where $\psi(x)$  and
$\psi_{\rm out}(x)$  denote local  composite operators  and asymptotic
fields for $\Lambda$, $\Sigma$ and $\Xi$.
Note  that, to  obtain  the  $\Xi\Lambda$-$\Xi\Sigma$ coupled  channel
potentials,  we need  Z  factors  for $\Lambda$  and  $\Sigma$ in  the
combination $\sqrt{Z_{\Lambda}/Z_{\Sigma}}$.
For  point-wall correlators,  the  plateau regions  are identified  as
$15-20$  for $\Lambda$  and $\Sigma$,  and $20-25$  for $\Xi$.   Since
wall-wall correlators are too noisy  to determine the plateau regions,
we employ two regions (i) 10-15  and (ii) 15-20.  Since the results do
not   change  so   much,   i.e.,   the  fit   with   10-15  leads   to
$\sqrt{Z_{\Lambda}/Z_{\Sigma}} =  1.02(3)$ whereas the one  with 15-20
leads  to  $\sqrt{Z_{\Lambda}/Z_{\Sigma}}  = 1.05(9)$,  we  adopt  the
result with 10-15.  The results of  the hyperon masses are as follows:
$m_{\Lambda} = 1.121(3)$ GeV, $m_{\Sigma} = 1.204(3)$ GeV and $m_{\Xi}
= 1.336(1)$ GeV.

\section{$\Xi\Sigma$ single channel for $I=3/2$ sector}

To   obtain  the   $\Xi\Sigma$  potentials   ($I=3/2$),  we   use  the
time-dependent HAL QCD method \cite{HALQCD:2012aa}.
For this purpose, we define the R-correlator for $\Xi\Sigma$ as
\begin{equation}
  R_{\Xi\Sigma}(\vec x-\vec y,t)
  \equiv
  e^{+(m_{\Xi}+m_{\Sigma})t}
  \left\langle
  0
  \left|
  T\left[
    \Xi(\vec x,t)
    \Sigma(\vec y,t)
    \cdot
    \mathcal{J}_{\Xi\Sigma}(t=0)
    \right]
  \right|
  0
  \right\rangle,
  \label{eq:R-corr1}
\end{equation}
where $\mathcal{J}_{\Xi\Sigma}$ denotes a wall source for $\Xi\Sigma$.
R-correlator satisfies the time-dependent Schr\"odinger-like equation,
which involves a fourth order time derivative \cite{Ishii:2016zsf}.
The $\Xi\Sigma$  potential for  $I=3/2$ should  be obtained  from this
equation. However, since the numerical  evaluation of the fourth order
time  derivative is  not  stable yet,  we  solve its  non-relativistic
approximation keeping the leading order of the derivative expansion of
the non-local potential as
\begin{equation}
  \left(
  -\frac{\partial}{\partial t}
  + \frac{\nabla^2}{2\mu}
  \right)
  R_{\Xi\Sigma}(\vec r,t)
  =
  V_{\Xi\Sigma}(\vec r)
  R_{\Xi\Sigma}(\vec r,t),
\end{equation}
where $\mu  \equiv 1/(1/m_{\Xi}  + 1/m_{\Sigma})$ denotes  the reduced
mass.

\begin{figure}[h]
  \begin{center}
    \includegraphics[angle=-90,width=\Figwidth]{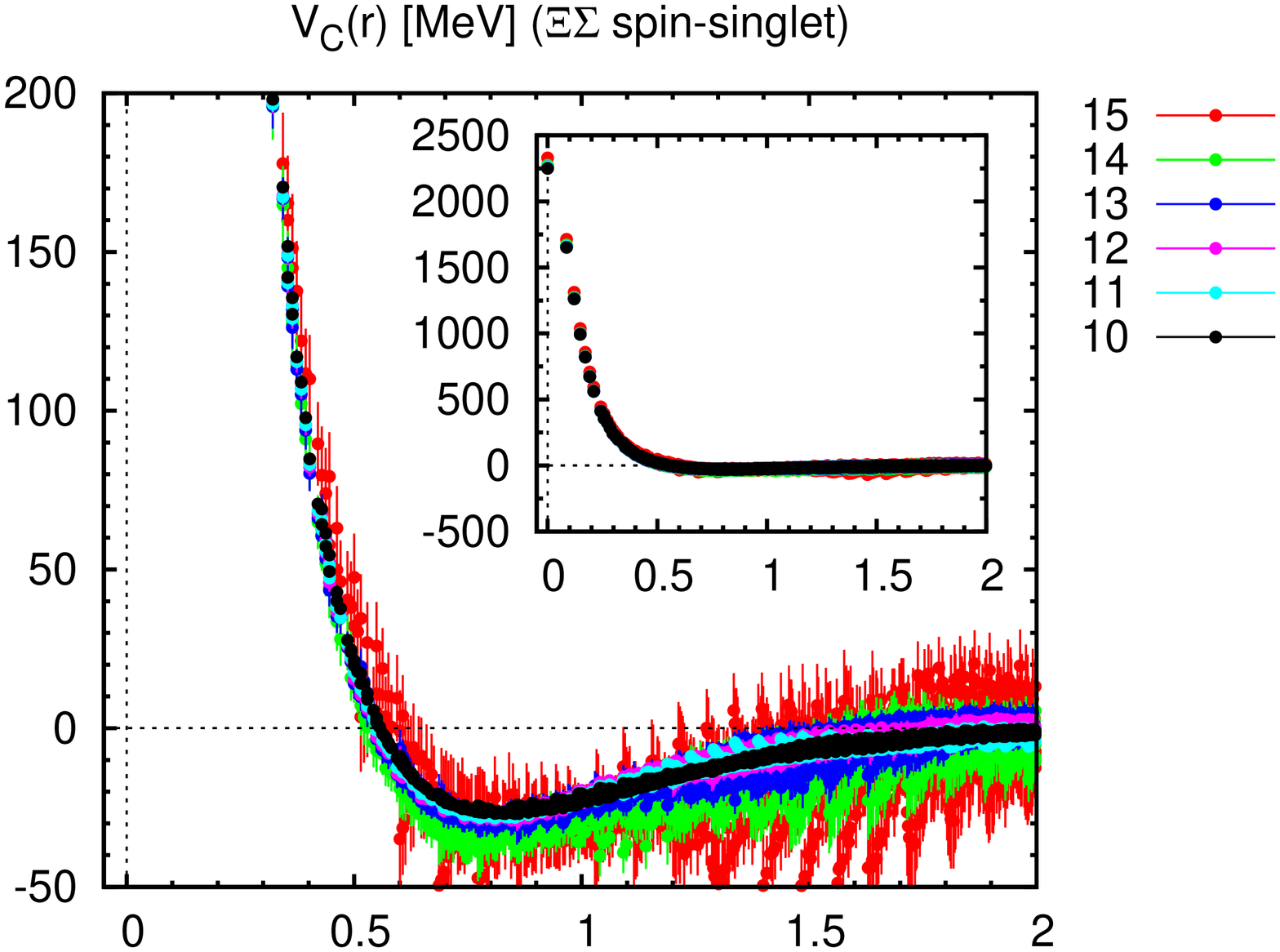}
    \\
    \includegraphics[angle=-90,width=\Figwidth]{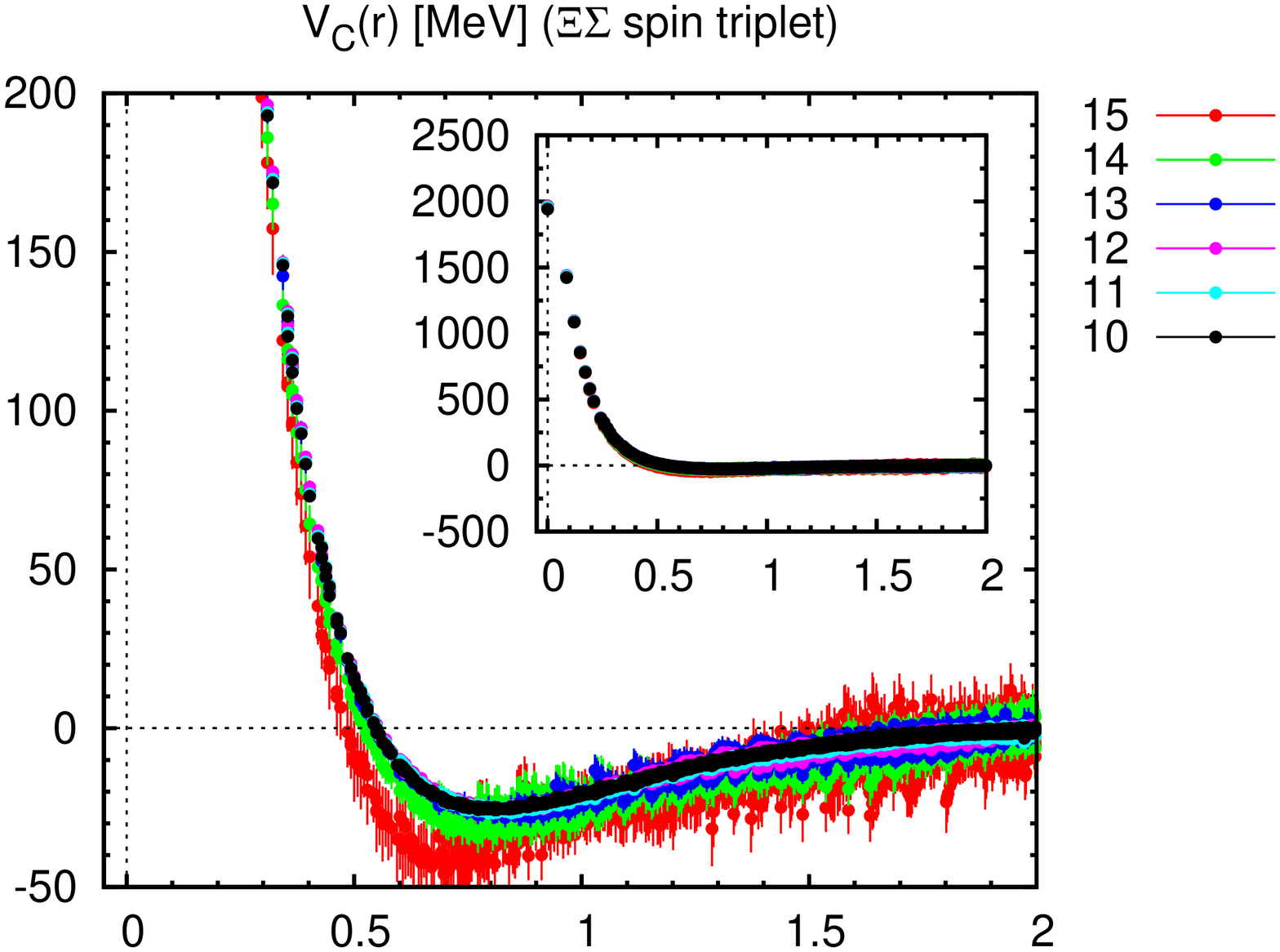}
    \includegraphics[angle=-90,width=\Figwidth]{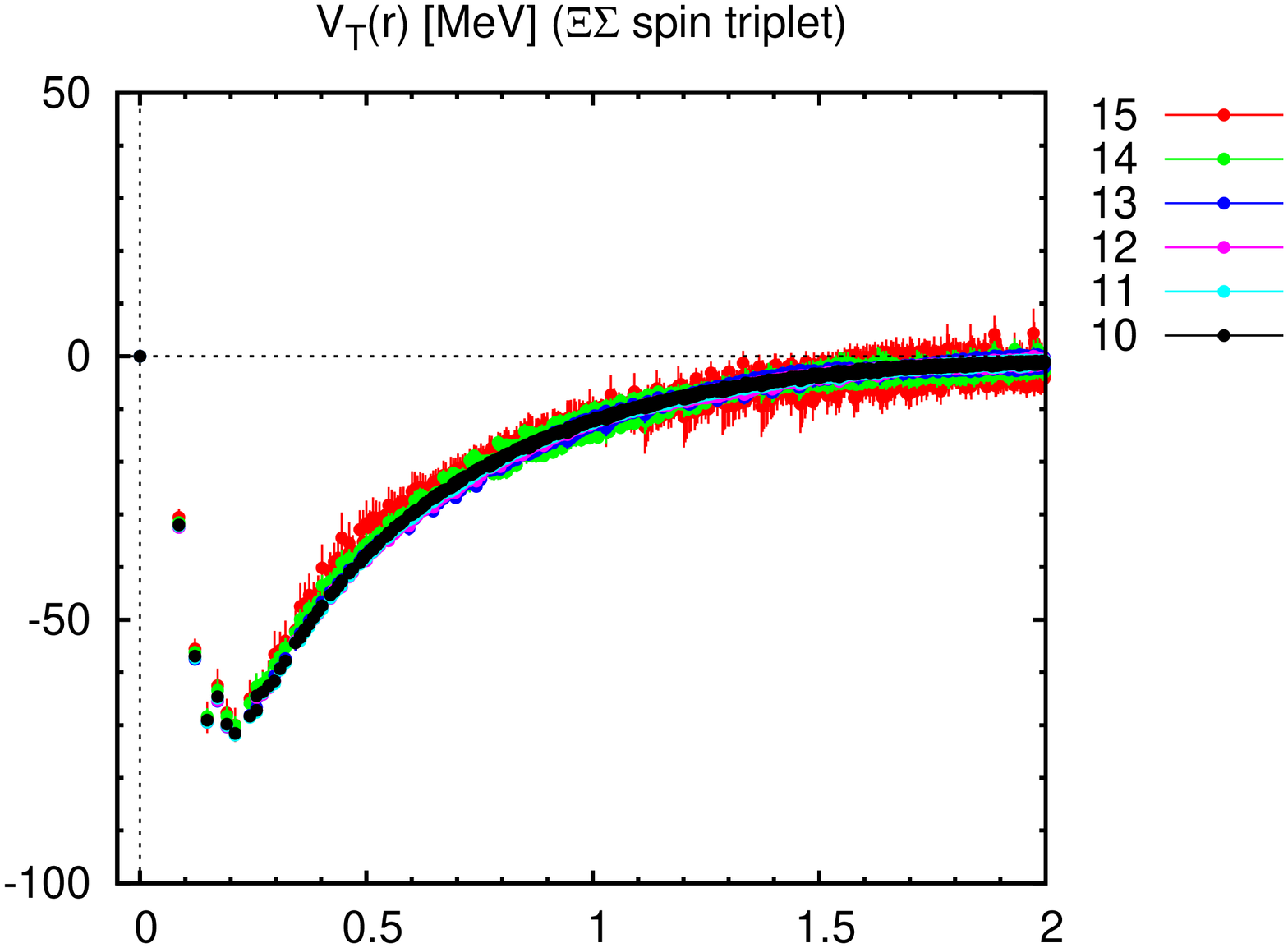}
  \end{center}
  \caption{$\Xi\Sigma$($I=3/2$) potentials. (a)  the central potential
    for   spin-singlet  sector,   (b)   the   central  potential   for
    spin-triplet  sector, (c)  the tensor  potential for  spin-triplet
    sector.}
  \label{fig:potential32}.
\end{figure}
\Fig{fig:potential32}(a)    shows    the    central    potential    of
$\Xi\Sigma$($I=3/2$) for spin-singlet sector obtained in the region $t
= 10-15$.
Since $t$-dependence  is seen to be  quite mild, we regard  that rough
convergence is achieved.
We see  that there  is a  repulsive core at  short distance,  which is
surrounded by  an attraction.  Qualitative  behavior is similar  to NN
case,  which   is  because  they   belong  to  the   same  irreducible
representation (irrep.)  ${\bf 27}$ in the flavor SU(3) limit.
\Fig{fig:potential32}(b)  and  (c) show  the  central  and the  tensor
potentials of $\Xi\Sigma$($I=3/2$) for spin-triplet sector obtained in
the region $t=10 - 15$.
Again, from the mild $t$-dependence,  we regard that rough convergence
is achieved.
We see that qualitative behaviors  are similar to NN potentials, i.e.,
the central potential has a repulsive  core, which is surrounded by an
attraction.
This is  because spin-triplet channel of  $\Xi\Sigma$($I=3/2$) belongs
to  the irrep.   ${\bf  10^*}$ in  the flavor  SU(3)  limit, to  which
spin-triple sector of NN belongs.

In  obtaining  the  $\Xi\Sigma$   potential,  we  replace  the  factor
$e^{(m_\Xi + m_\Sigma)t}$ in \Eq{eq:R-corr1} by a product of two-point
correlators(point-wall) of  $\Xi$ and $\Sigma$ so  that the correlated
statistical noises may cancel.  Therefore, although the $t$-dependence
of the  potential is mild, $t$  should be large enough  to achieve the
ground  state  saturation  of   two-point  correlators  of  $\Xi$  and
$\Sigma$, i.e., $t \agt 20$.

\begin{figure}[h]
\begin{center}
  \includegraphics[angle=-90,width=\Figwidth]{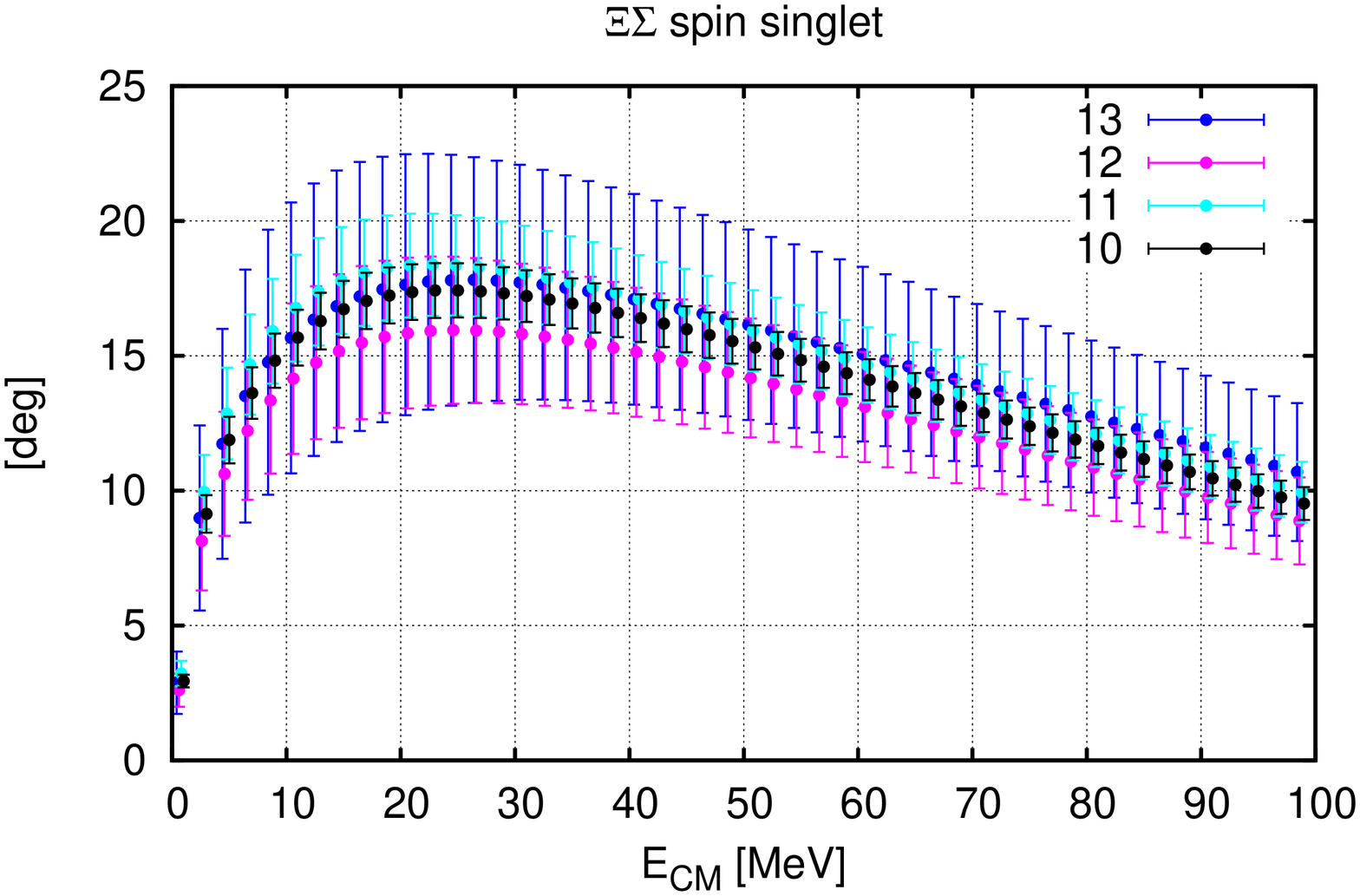}
  \includegraphics[angle=-90,width=\Figwidth]{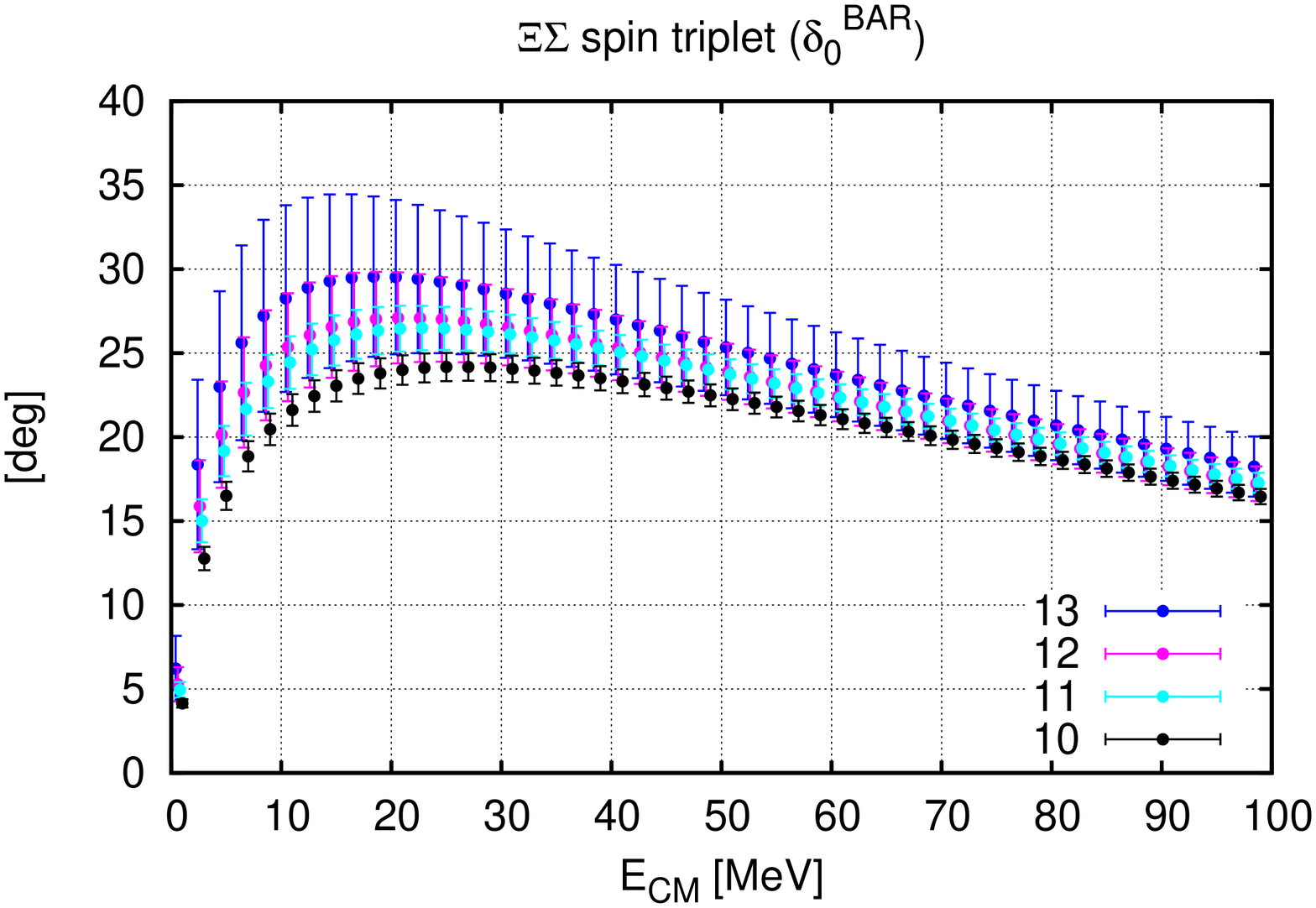}
  \\
  \includegraphics[angle=-90,width=\Figwidth]{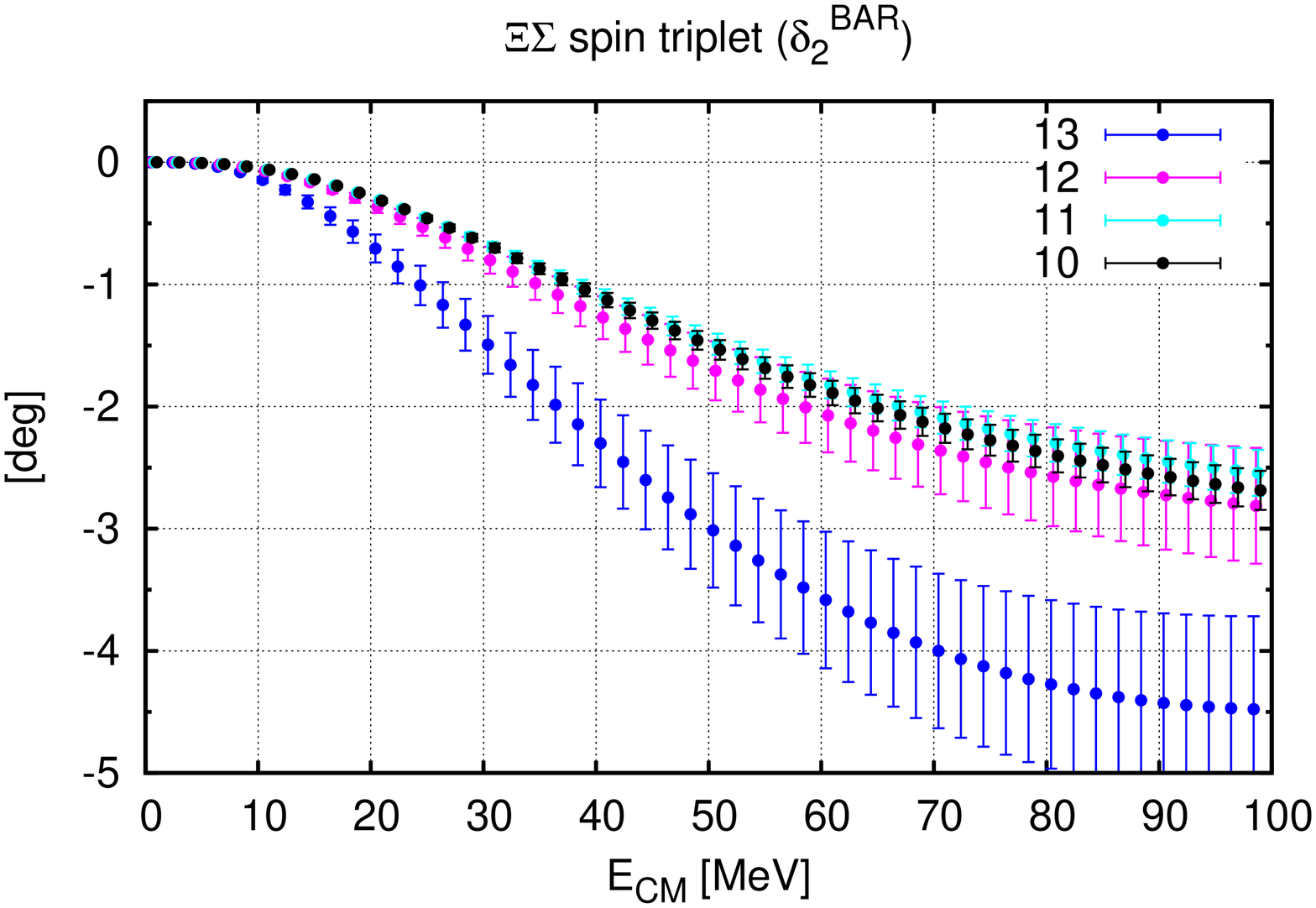}
  \includegraphics[angle=-90,width=\Figwidth]{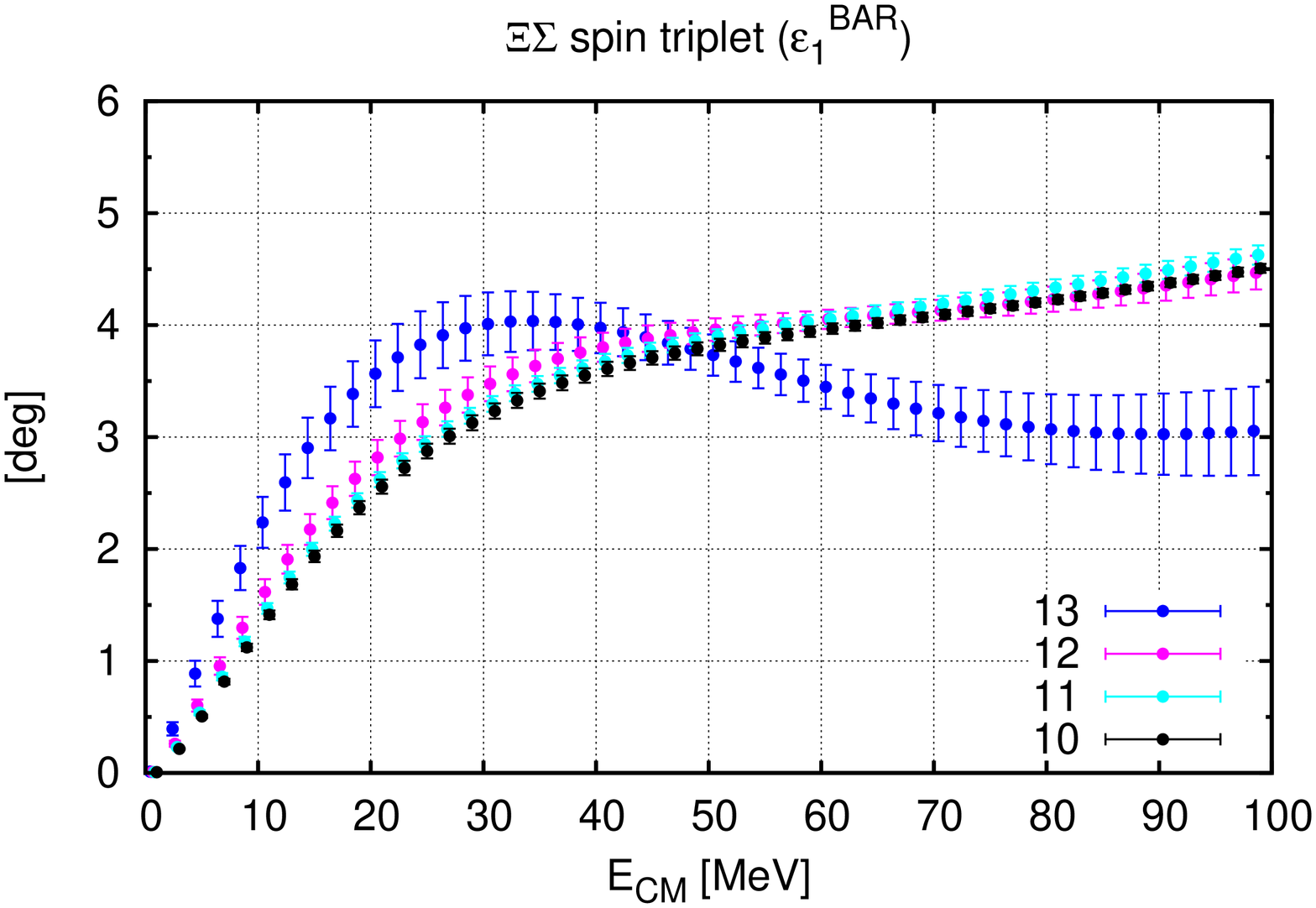}
\end{center}
  \caption{The  scattering   phase  shift   from  $\Xi\Sigma$($I=3/2$)
    potentials  given  in  Fig.~\protect\ref{fig:potential32}  in  the
    region  $t=10-13$.  (a)  $^1S_0$  phase shift,  (b) $^3S_1$  phase
    shift $\bar\delta_0$, (c) $^3D_1$  phase shift $\bar\delta_2$, (d)
    mixing parameter $\bar\epsilon_1$.}
  \label{fig:phaseshift}
\end{figure}

We use  these potentials in  the Schr\"odinger equation to  obtain the
scattering phase shifts.
We first perform a fit to  obtain smooth potentials, which are used to
solve the Schr\"odinger equation.
%
\Fig{fig:phaseshift} shows  the scattering  phase shift  obtained from
the potentials in the region $t=10-13$.
We see  that qualitative behaviors are  similar to the NN  phase shift
except for the two points (i) the  strength is weak, and (ii) there is
no bound state in spin-triplet sector.

The   phase   shift   $\bar\delta_2$    and   the   mixing   parameter
$\bar\epsilon_1$ at $t=13$  deviate from the others.   This is because
of  a technical  reason that  the fit  of the  potential falls  into a
slightly  different  minimum.   Improvement  of  the  statistics  will
resolve it.

\section{$\Xi\Lambda$-$\Xi\Sigma$ coupled channel for $I=1/2$ sector}

To  obtain  the  $\Xi\Lambda$-$\Xi\Sigma$  coupled  channel  potential
($I=1/2$),  we   use  coupled  channel  extension   of  time-dependent
Schr\"odinger-like equation.
We define the R-correlators
\begin{eqnarray}
  R_{\Xi\Lambda}(\vec x-\vec y,t; \mathcal{J})
  &\equiv&
  e^{+(m_{\Xi} + m_{\Lambda})t}
  \left\langle 0 \left|
  T\left[
    \Xi(\vec x, t)\Lambda(\vec y,t)\cdot \mathcal{J}(t=0)
    \right]
  \right| 0 \right\rangle
  \\\
  R_{\Xi\Sigma}(\vec x - \vec y,t; \mathcal{J})
  &\equiv&
  e^{+(m_{\Xi} + m_{\Sigma})t}
  \left\langle 0 \left|
  T\left[
    \Xi(\vec x, t)\Sigma(\vec y,t)\cdot \mathcal{J}(t=0)
    \right]
  \right| 0 \right\rangle,
  \label{eq:R-corr2}
\end{eqnarray}
where  $\mathcal{J}=\mathcal{J}_{\Xi\Lambda}, \mathcal{J}_{\Xi\Sigma}$
denote   the   wall   sources  for   $\Xi\Lambda$   and   $\Xi\Sigma$,
respectively.
The coupled channel extension of the time-dependent Schr\"odinger-like
equation involves  fourth time derivative  \cite{Ishii:2016zsf}. Since
the numerical evaluation of fourth  time derivative is still unstable,
we solve  its non-relativistic approximation keeping  only the leading
order of the derivative expansion of the non-local potentials as
\begin{equation}
  \left[
    \begin{array}{c}
      \left(
      - \frac{\partial}{\partial t}
      + \frac{\nabla^2}{2\mu_{\Xi\Lambda}}
      \right)
      R_{\Xi\Lambda}(\vec r,t; \mathcal{J})
      \\
      \left(
      - \frac{\partial}{\partial t}
      + \frac{\nabla^2}{2\mu_{\Xi\Sigma}}
      \right)
      R_{\Xi\Sigma}(\vec r,t; \mathcal{J})
    \end{array}
    \right]
  =
  \left[
    \begin{array}{cc}
      V_{\Xi\Lambda;\Xi\Lambda}(\vec r)
      &
      \zeta_0
      \zeta^{+t/a}
      V_{\Xi\Lambda;\Xi\Sigma}(\vec r)
      \\
      \zeta_0^{-1}
      \zeta^{-t/a}
      V_{\Xi\Sigma;\Xi\Lambda}(\vec r)
      &
      V_{\Xi\Sigma;\Xi\Sigma}(\vec r)
    \end{array}
    \right]
  \cdot
  \left[
    \begin{array}{c}
      R_{\Xi\Lambda}(\vec r,t; \mathcal{J})
      \\
      R_{\Xi\Sigma}(\vec r,t; \mathcal{J})
    \end{array}
    \right],
  \label{eq:coupled-channel}
\end{equation}
where $\zeta \equiv e^{(m_\Sigma - m_{\Lambda})a}$ and $\zeta_0 \equiv
\sqrt{Z_{\Lambda}/Z_{\Sigma}}$  with  $Z_{\Lambda}$  and  $Z_{\Sigma}$
being the Z factors for the local composite operators of $\Lambda$ and
$\Sigma$, respectively.
$\mu_{\Xi\Lambda}   \equiv   \frac1{1/m_\Xi    +   1/m_\Lambda}$   and
$\mu_{\Xi\Sigma}  \equiv  \frac1{1/m_\Xi  +  1/m_\Sigma}$  denote  the
reduced masses for $\Xi\Lambda$ and $\Xi\Sigma$, respectively.
Note   that   four    unknowns   $V_{\Xi\Lambda;\Xi\Lambda}(\vec   r)$,
$V_{\Xi\Lambda;\Xi\Sigma}(\vec r)$, $V_{\Xi\Sigma;\Xi\Lambda}(\vec r)$
and  $V_{\Xi\Sigma;\Xi\Sigma}(\vec   r)$  are  determined   from  four
equations        in         Eq.~(\ref{eq:coupled-channel})        with
$\mathcal{J}=\mathcal{J}_{\Xi\Lambda}$ and $\mathcal{J}_{\Xi\Sigma}$.

\begin{figure}[h]
  \begin{center}
    \includegraphics[angle=-90,width=\Figwidth]{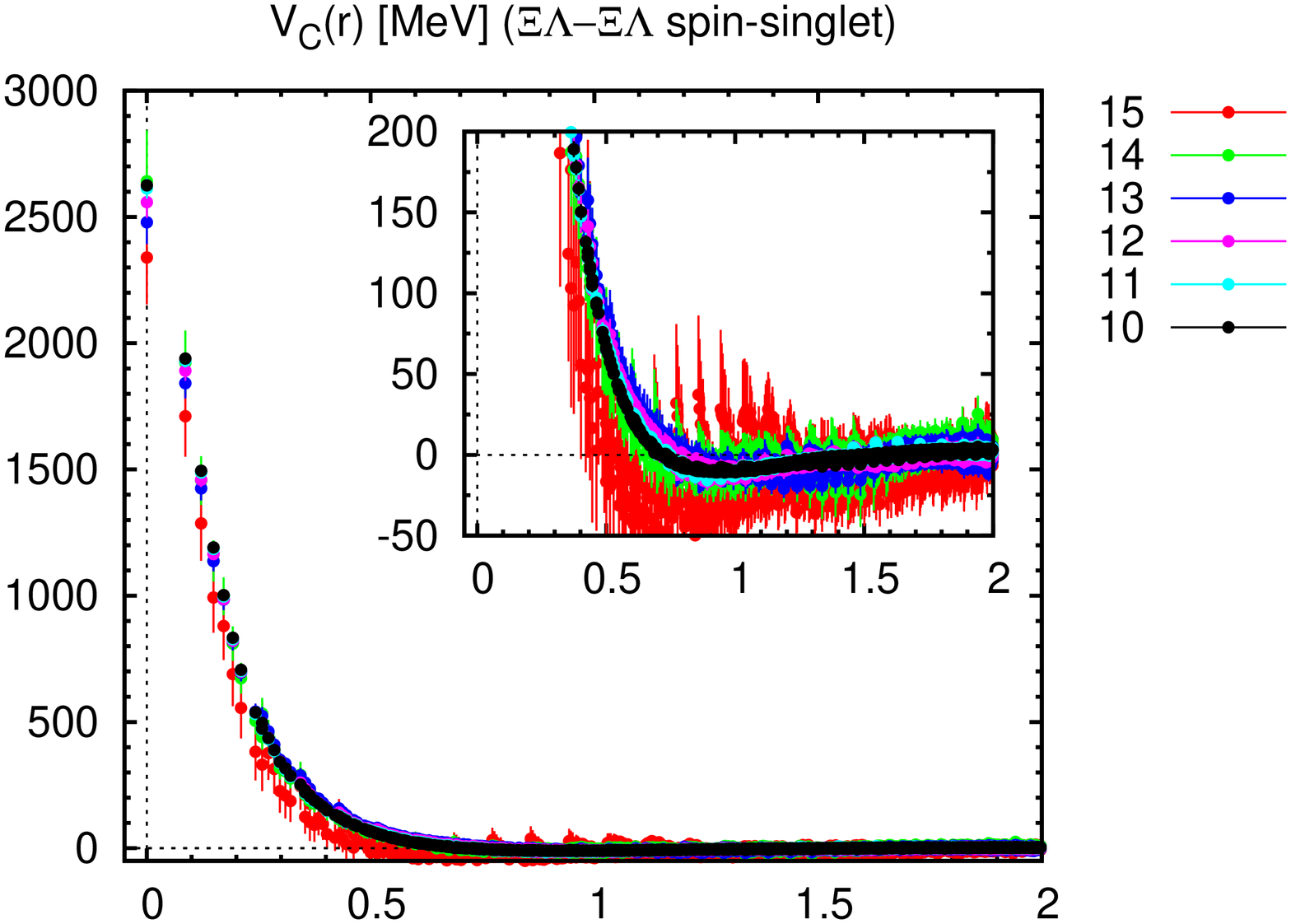}
    \includegraphics[angle=-90,width=\Figwidth]{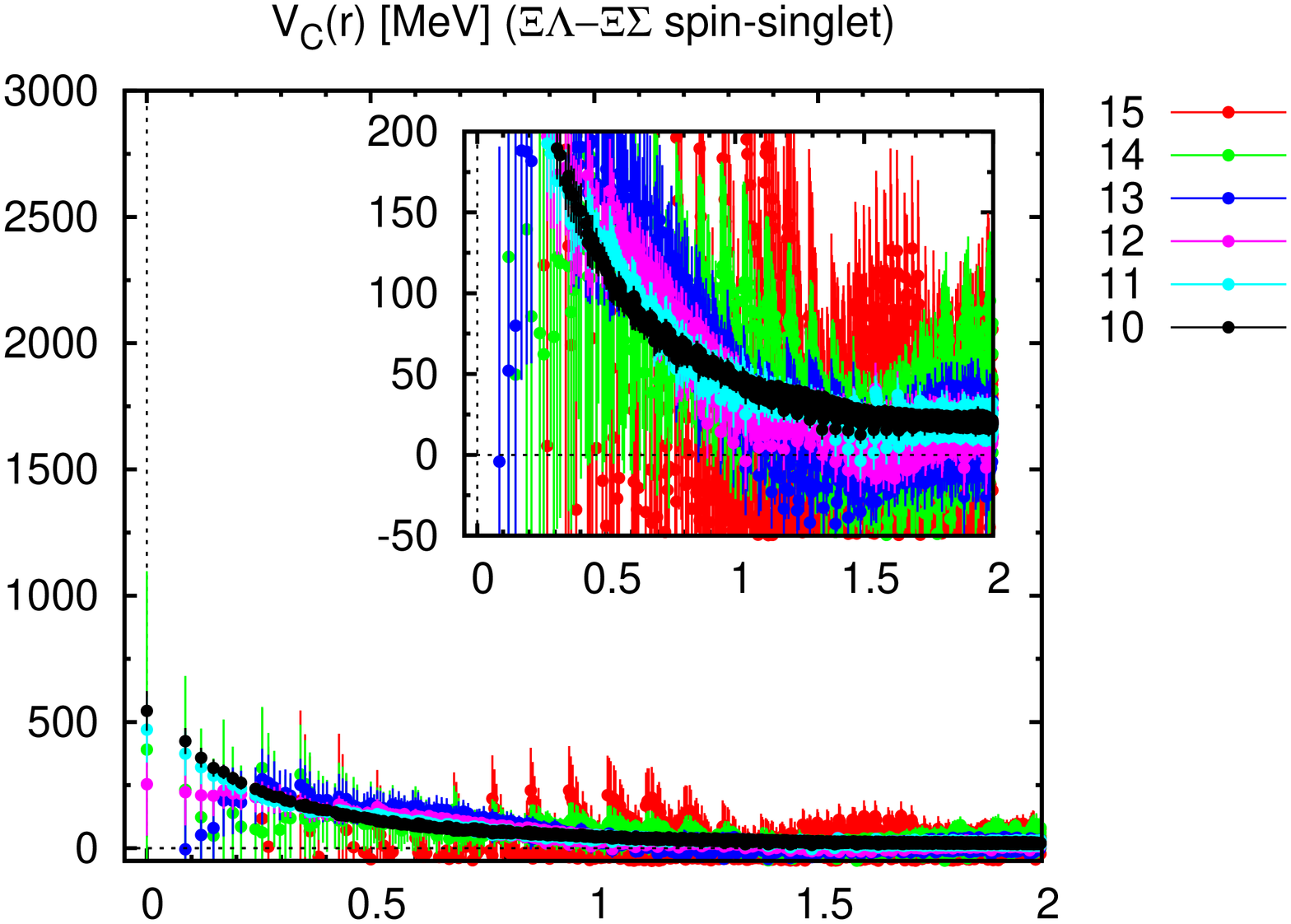}
    \\
    \includegraphics[angle=-90,width=\Figwidth]{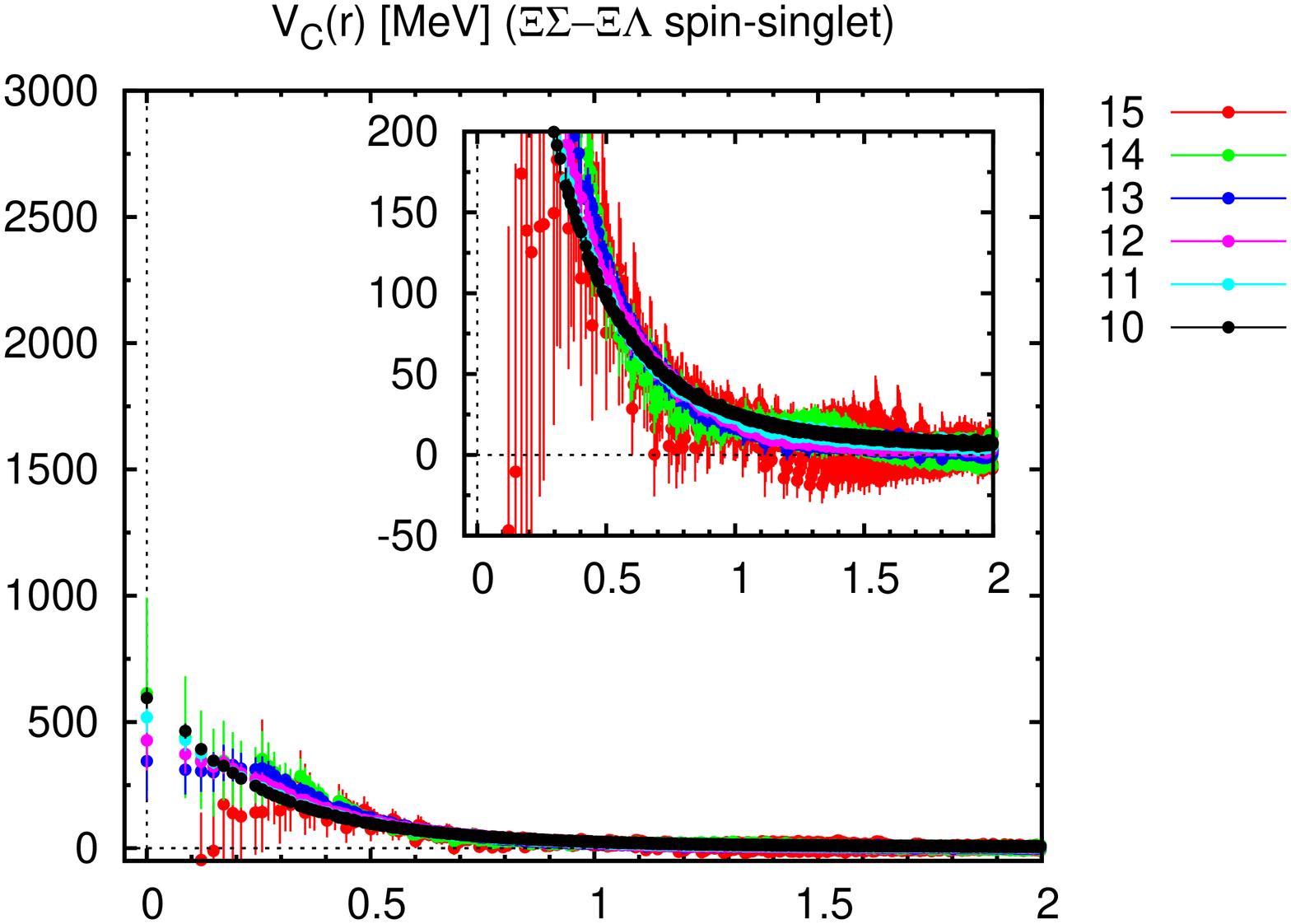}
    \includegraphics[angle=-90,width=\Figwidth]{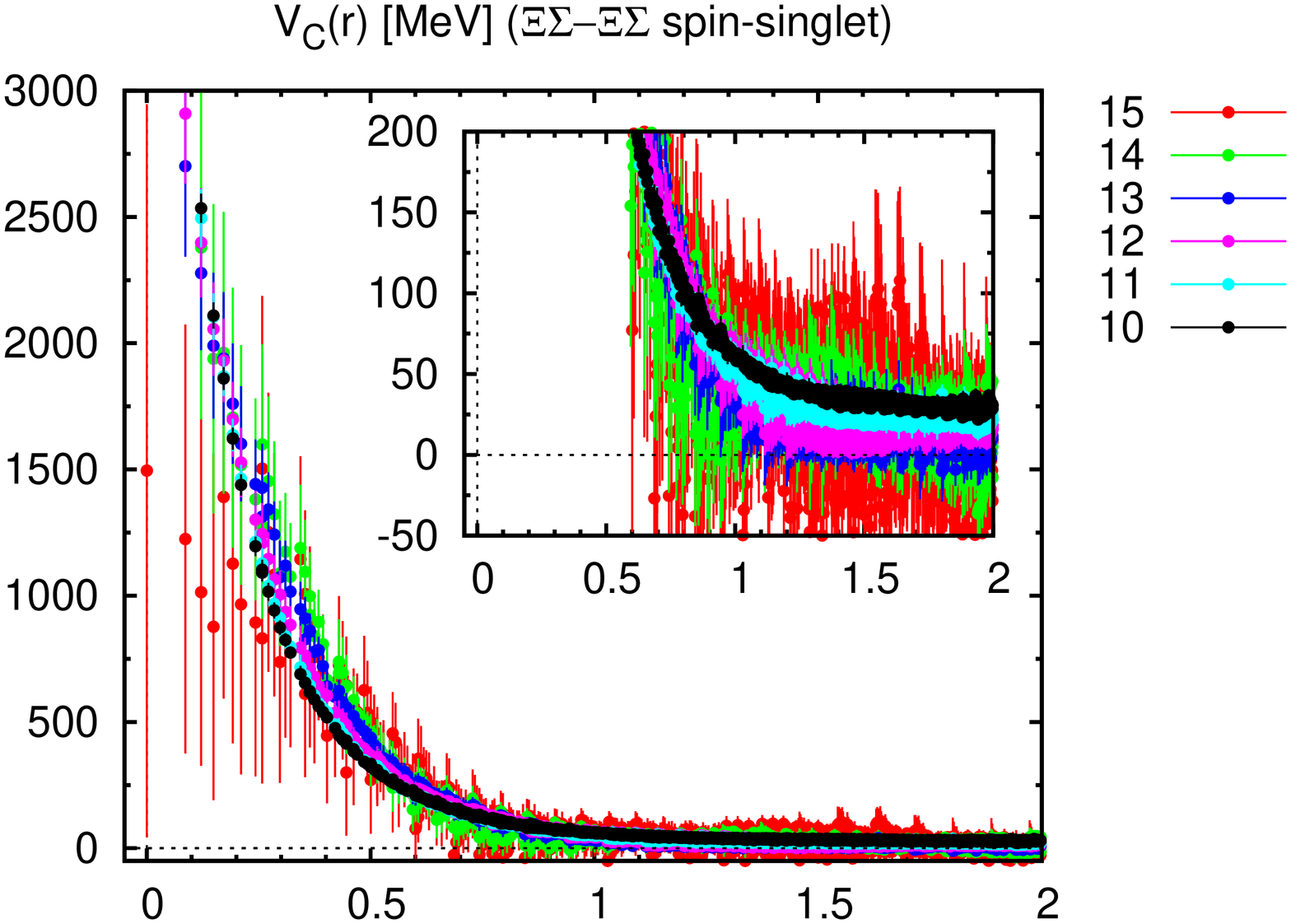}
  \end{center}
  \caption{The  $\Xi\Lambda$-$\Xi\Sigma$   ($I=1/2$)  coupled  channel
    central potentials in singlet spin channel.}
  \label{fig:potential121}
\end{figure}
\Fig{fig:potential121}  shows   the  $\Xi\Lambda$-$\Xi\Sigma$  coupled
channel  central potentials  for spin-singlet  sector obtained  in the
region $t = 10-15$.
We see that they are noisy, which is often the case if it contains the
irrep.  ${\bf 8_{\rm S}}$ in the flavor SU(3) limit.
$t$-dependence is seen  to be mild except that the  long distance part
of $\Xi\Lambda$-$\Xi\Sigma$ and $\Xi\Sigma$-$\Xi\Sigma$ potentials are
still changing.
Flavor SU(3) limit is helpful  to understand the qualitative behaviors
of these  potentials \cite{Inoue:2011ai}.  In the  flavor SU(3) limit,
these coupled channel potentials are  related to those for the irreps.
${\bf 27}$ and ${\bf 8_{\rm S}}$ as
$V_{\Xi\Lambda;\Xi\Lambda}  =  \frac{9}{10}  V^{\bf 27}  +  \frac1{10}
V^{\bf 8_{\rm S}}$,
$V_{\Xi\Lambda;\Xi\Sigma}  =  V_{\Xi\Sigma;\Xi\Lambda}  =  -\frac3{10}
  V^{\bf    27}    +    \frac3{10}     V^{\bf    8_{\rm    S}}$
and
$V_{\Xi\Sigma;\Xi\Sigma} = \frac1{10} V^{\bf 27} + \frac{9}{10} V^{\bf
  8_{\rm S}}$.
These relations  together with  the potentials  for the  irreps. ${\bf
  27}$  and   ${\bf  8_{\rm   S}}$  given   in  Ref\cite{Inoue:2011ai}
qualitatively explain (i) the existence of attractive pocket at medium
distance in $\Xi\Lambda$-$\Xi\Lambda$ potential,  and (ii) strength of
the repulsive cores of these  four coupled channel potentials at short
distance.

\begin{figure}[h]
  \begin{center}
    \includegraphics[angle=-90,width=\Figwidth]{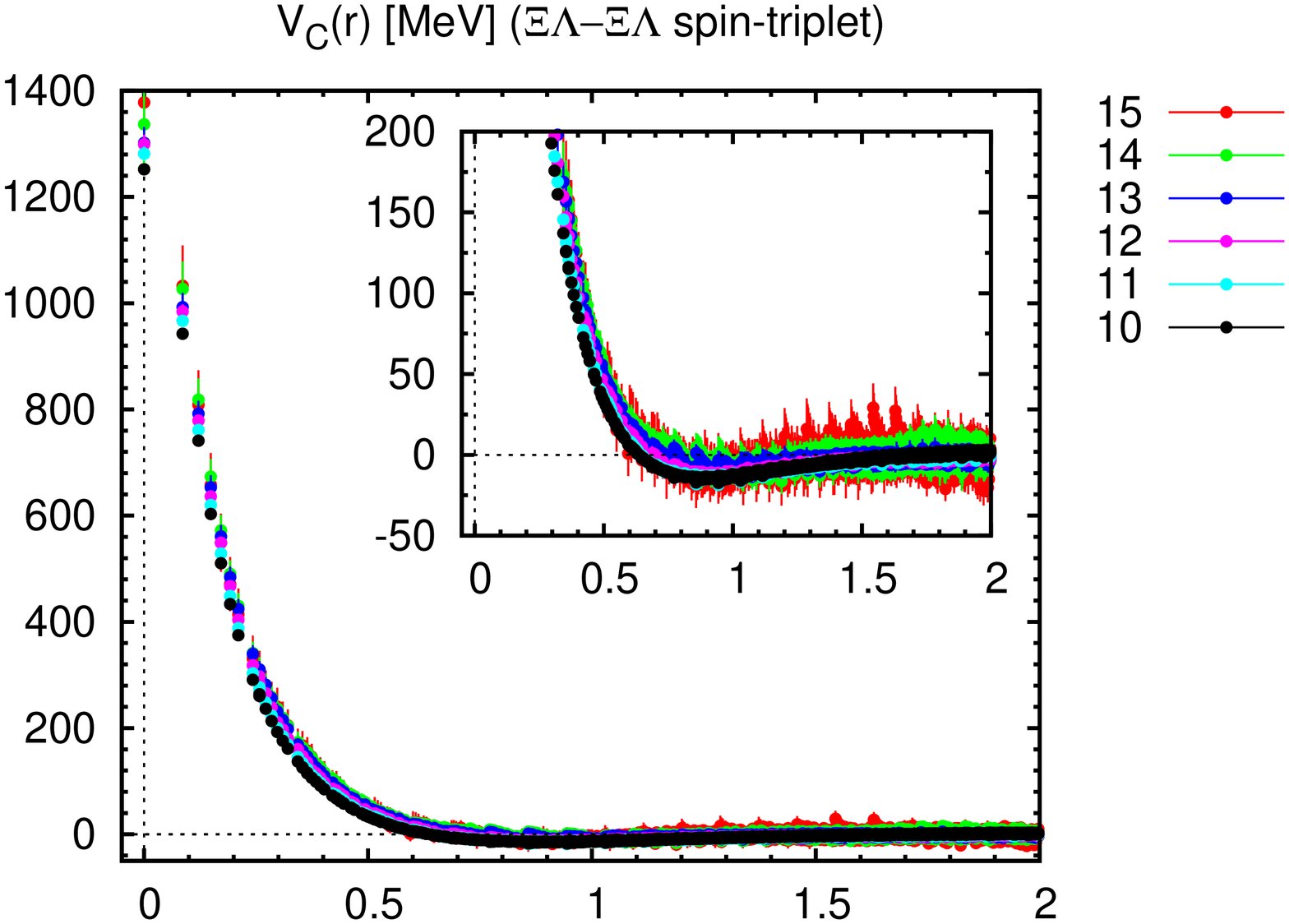}
    \includegraphics[angle=-90,width=\Figwidth]{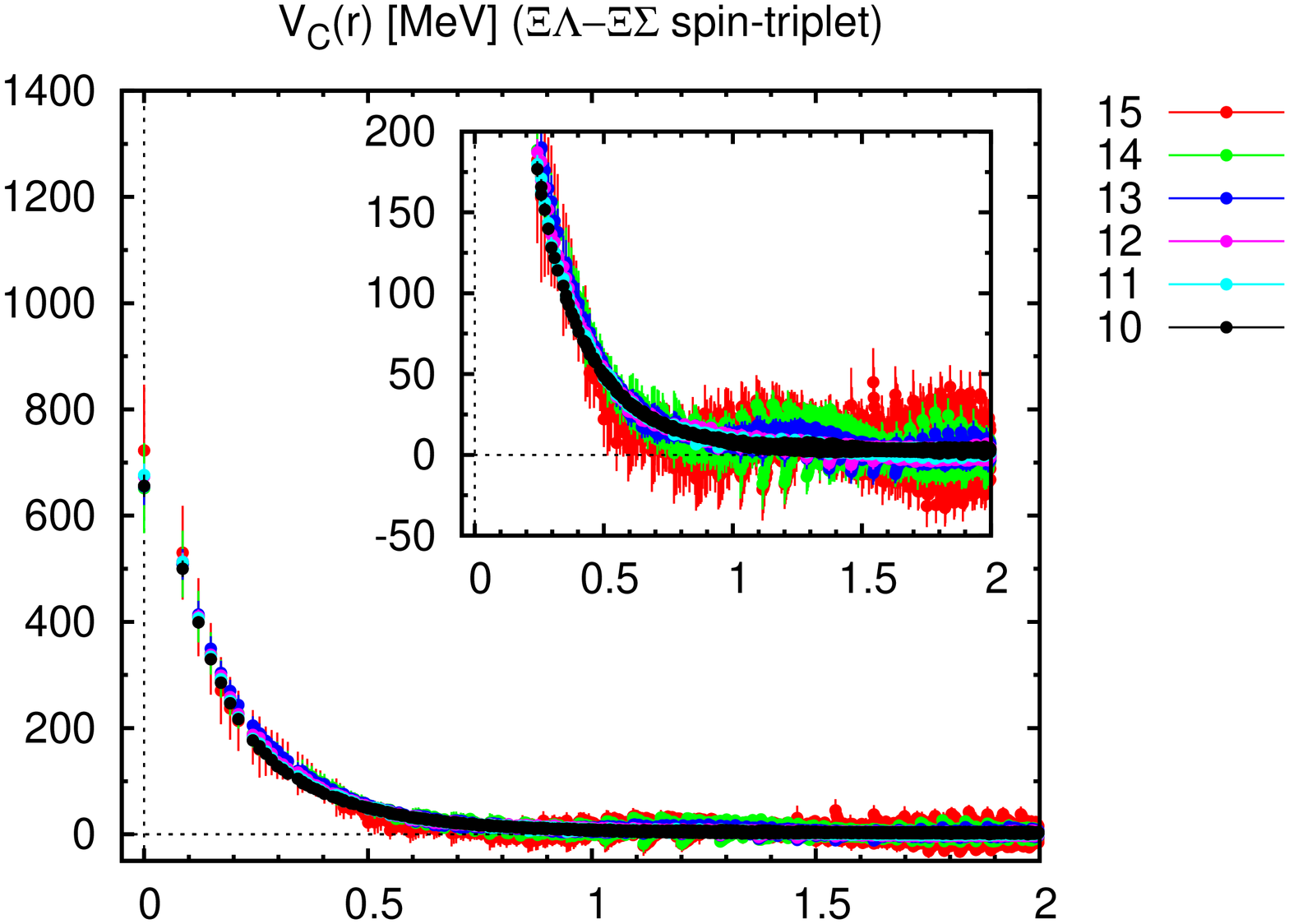}
    \\
    \includegraphics[angle=-90,width=\Figwidth]{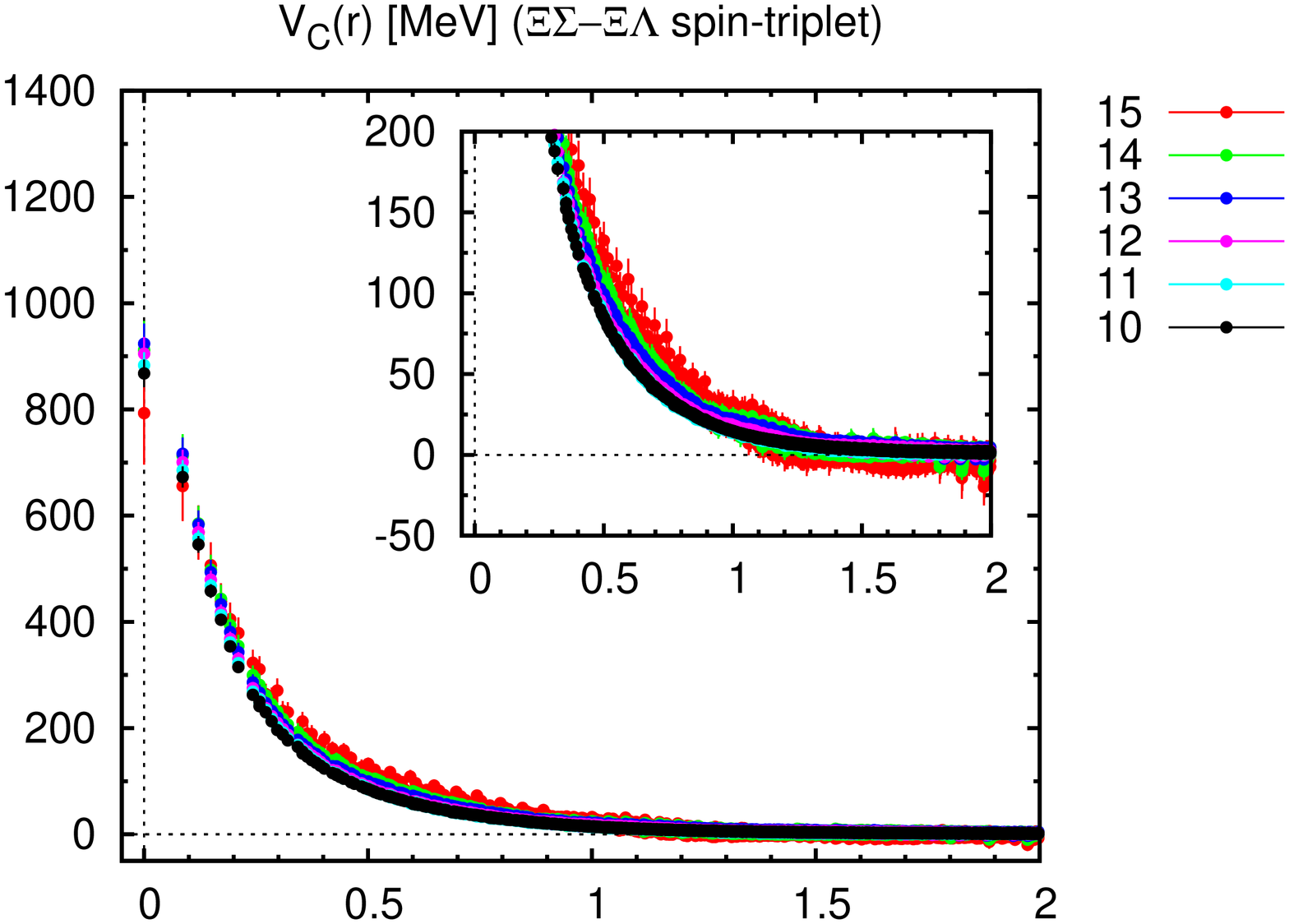}
    \includegraphics[angle=-90,width=\Figwidth]{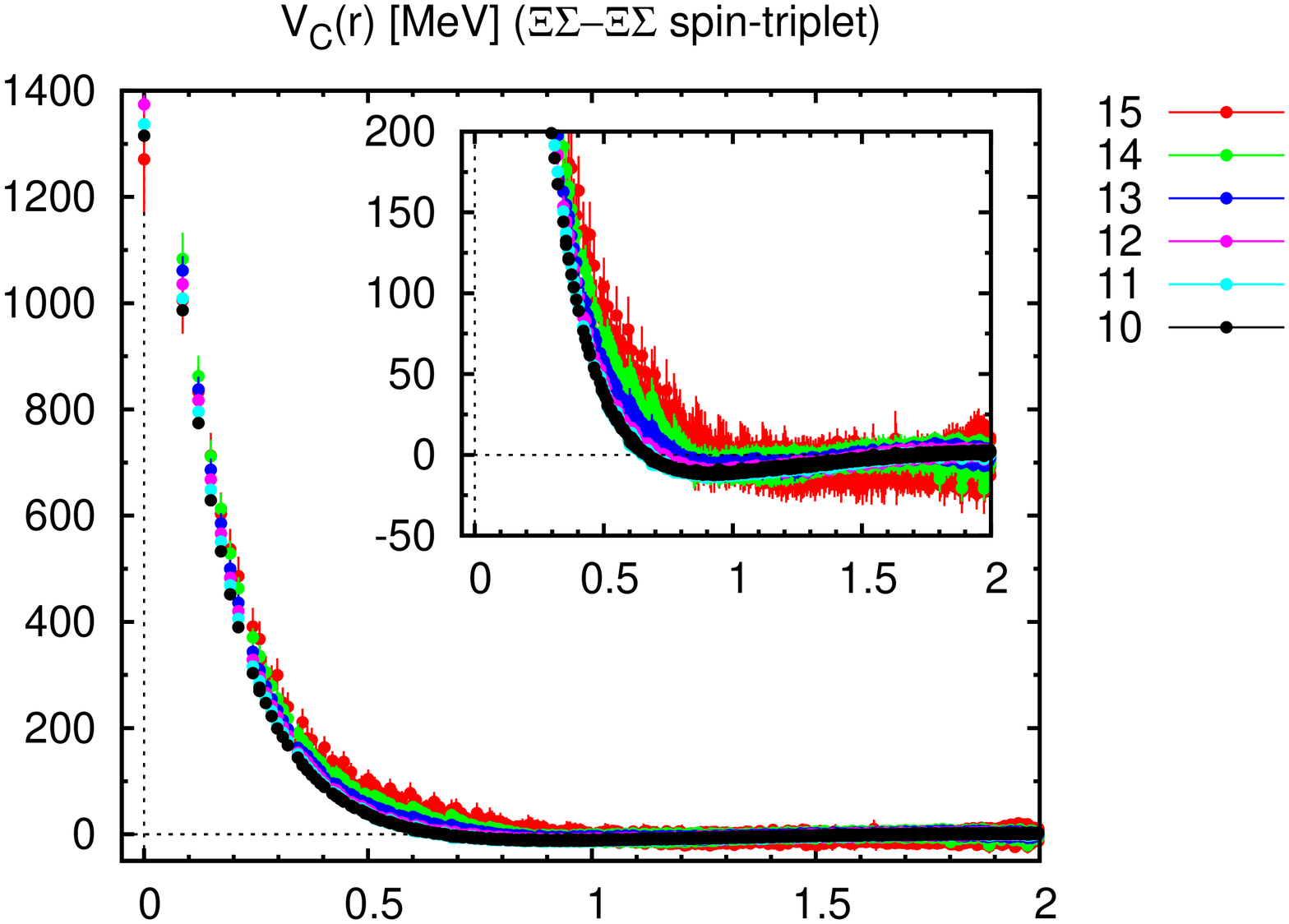}
  \end{center}
  \caption{The  $\Xi\Lambda$-$\Xi\Sigma$   ($I=1/2$)  coupled  channel
    central potentials in triplet spin channel.}
  \label{fig:potential123c}
\end{figure}
\begin{figure}[h]
  \begin{center}
    \includegraphics[angle=-90,width=\Figwidth]{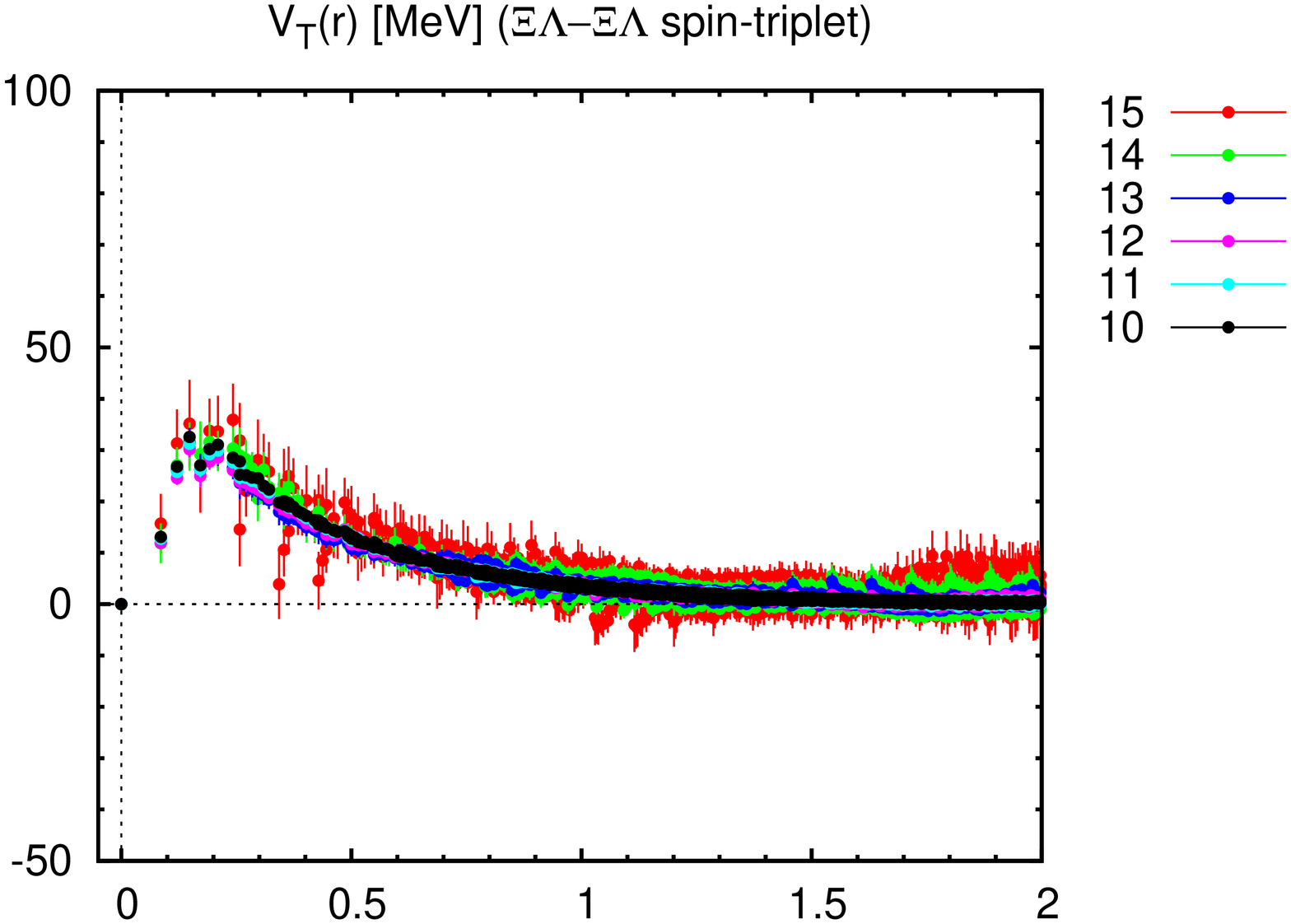}
    \includegraphics[angle=-90,width=\Figwidth]{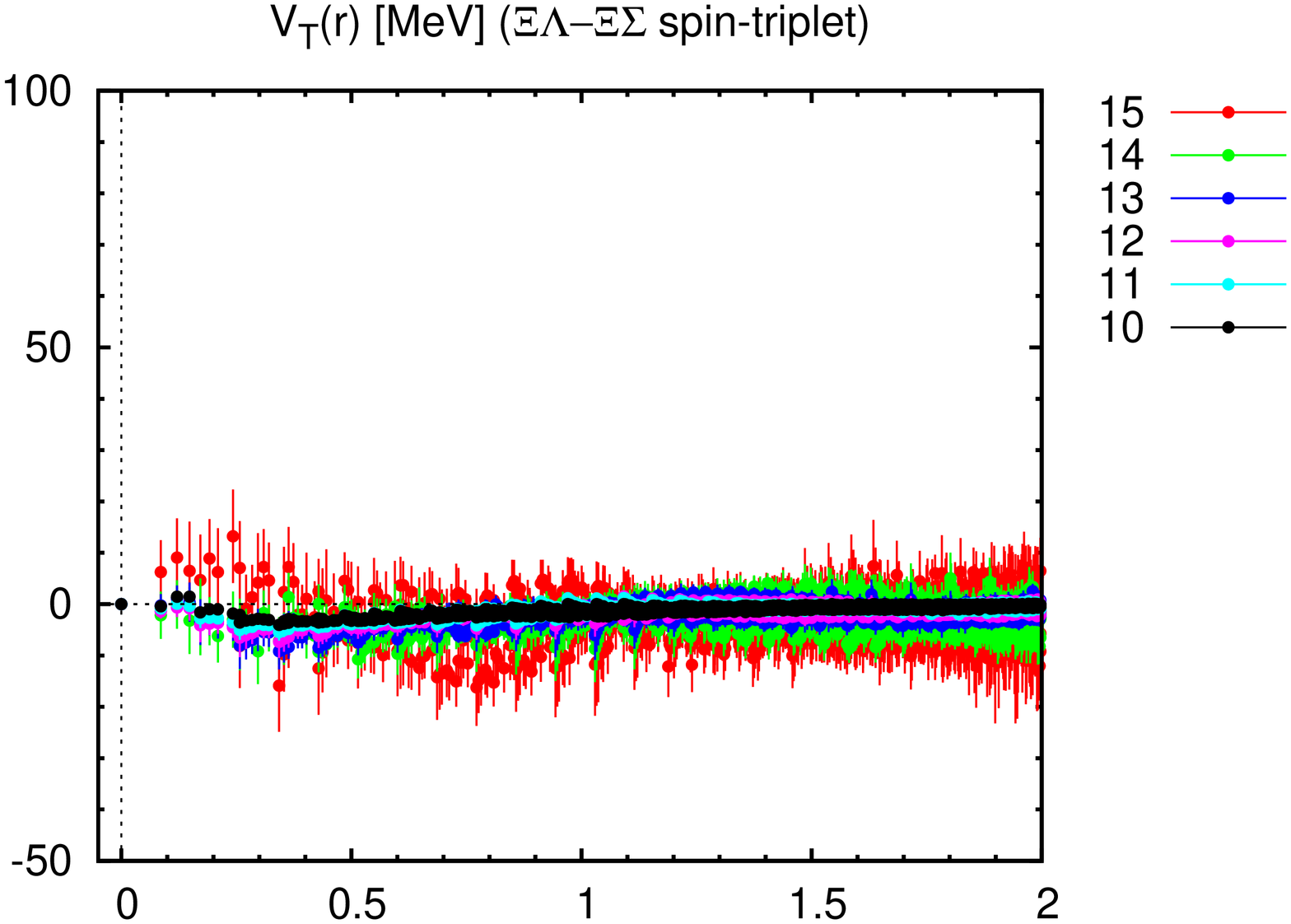}
    \\
    \includegraphics[angle=-90,width=\Figwidth]{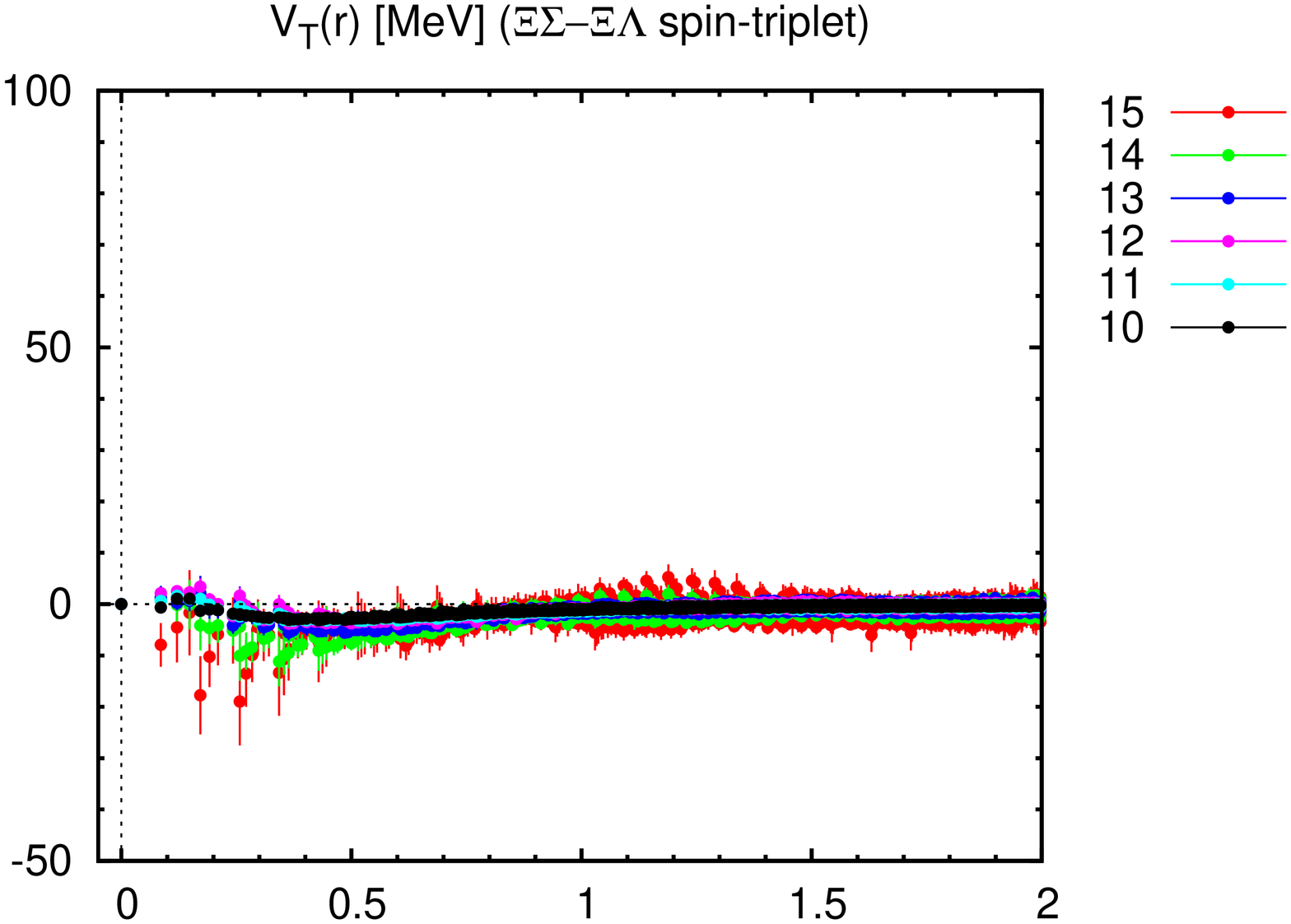}
    \includegraphics[angle=-90,width=\Figwidth]{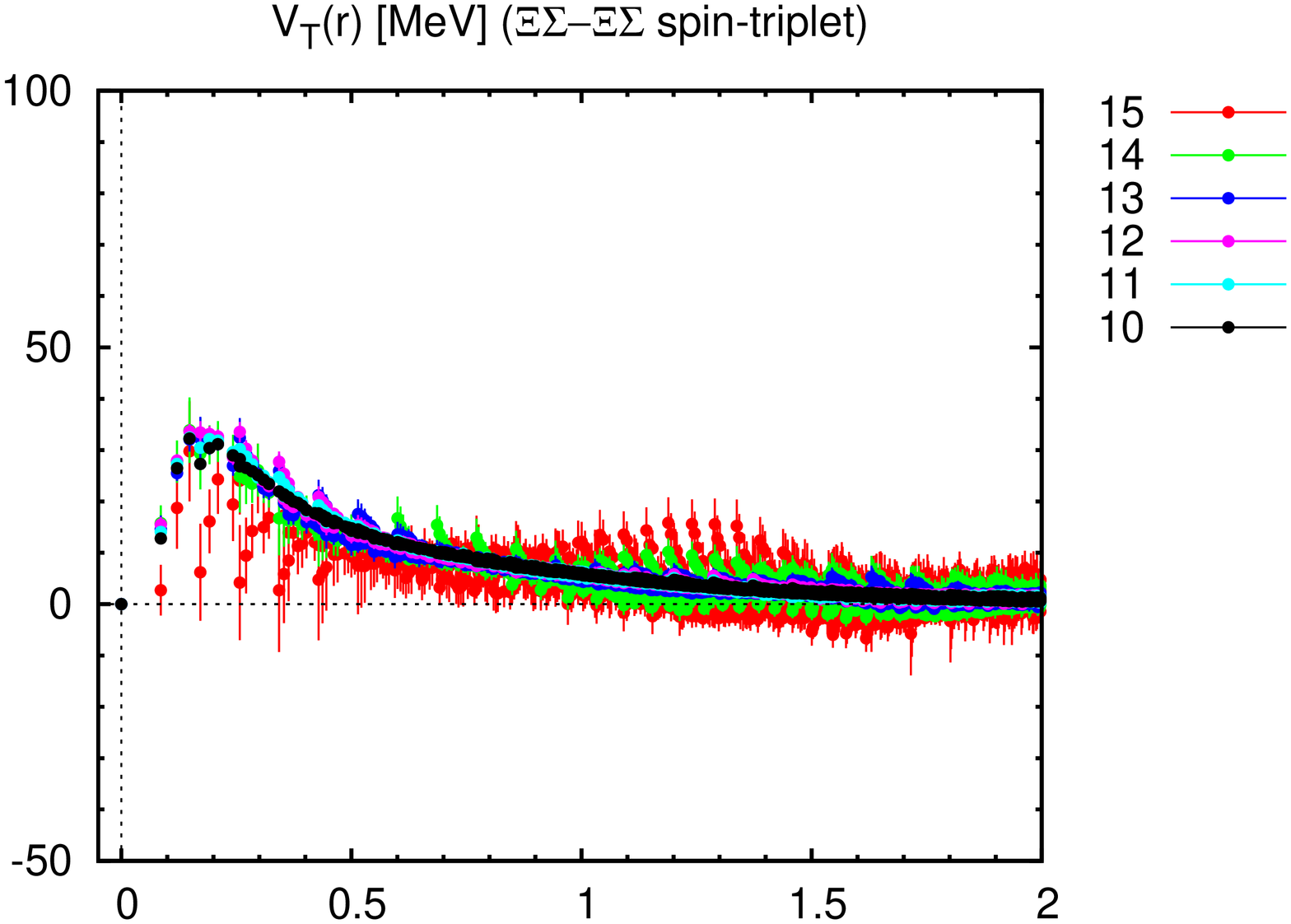}
  \end{center}
  \caption{The  $\Xi\Lambda$-$\Xi\Sigma$   ($I=1/2$)  coupled  channel
    tensor potentials in triplet spin channel.}
  \label{fig:potential123t}
\end{figure}
\Fig{fig:potential123c}    and   \Fig{fig:potential123t}    show   the
$\Xi\Lambda$-$\Xi\Sigma$   coupled   channel    central   and   tensor
potentials,  respectively, for  spin-triplet  sector  obtained in  the
region $t=10-15$.
$t$-dependences are  seen to be  mild. But the central  potentials for
the  $\Xi\Lambda$-$\Xi\Sigma$  and  $\Xi\Sigma$-$\Xi\Lambda$  fail  to
achieve the complete convergence at long distance.
To  understand their  behaviors qualitatively,  the potentials  in the
flavor SU(3) limit are helpful again. In the flavor SU(3) limit, these
coupled channel central and tensor  potentials are related to those of
the    irreps.     ${\bf   10}$    and    ${\bf    8_{\rm   A}}$    as
$V_{\Xi\Lambda;\Xi\Lambda}   =  V_{\Xi\Sigma;\Xi\Sigma}   =  \frac1{2}
V^{\bf     10}    +     \frac1{2}    V^{\bf     8_{\rm    A}}$     and
$V_{\Xi\Lambda;\Xi\Sigma}   =  V_{\Xi\Sigma;\Xi\Lambda}   =  \frac1{2}
V^{\bf  10}  -  \frac1{2}  V^{\bf  8_{\rm  A}}$,  which  explains
(i) existence of the weak  attractive pocket in the central potentials
of  $\Xi\Lambda$-$\Xi\Lambda$  and $\Xi\Sigma$-$\Xi\Sigma$  at  medium
distance,
(ii)  similarity between  the potentials  of $\Xi\Lambda$-$\Xi\Lambda$
and $\Xi\Sigma$-$\Xi\Sigma$  both for  central and  tensor potentials,
and
(iii)   weak  tensor   potentials   of  $\Xi\Lambda$-$\Xi\Sigma$   and
$\Xi\Sigma$-$\Xi\Lambda$.

Due to the similar reason as before, we replace the factors $e^{(m_\Xi
  + m_\Lambda)t}$ and $e^{(m_\Xi  + m_\Sigma)t}$ in \Eq{eq:R-corr2} by
products of two-point correlators of $\Xi$, $\Lambda$ and $\Sigma$.
We  also  replace  the  factors  $\zeta^{\pm t/a}$  by  the  ratio  of
two-point correlators  of $\Lambda$  and $\Sigma$  due to  a technical
reason.
Therefore, although $t$-dependences of  these potentials are mild, $t$
should  be large  enough to  achieve the  ground state  saturations of
$\Xi$, $\Lambda$ and $\Sigma$, i.e., $t \agt 20$.

\section{Summary}
We have  presented our  results of  the hyperon-hyperon  potentials in
$S=-3$  sector by  using the  2+1 flavor  QCD gauge  configurations at
almost  the physical  point ($m_{\pi}  \simeq  146$ MeV)  on the  huge
spatial  volume $L\simeq  8.1$ fm  generated  on $96^4$  lattice by  K
computer at AICS.
To obtain the potentials, we have used non-relativistic approximations
of the time-dependent Schr\"odinger-like  equations. We have presented
$\Xi\Sigma$  potentials   for  $I=3/2$   and  $\Xi\Lambda$-$\Xi\Sigma$
coupled channel potentials for $I=1/2$.  We have seen that qualitative
behaviors are consistent with those in the flavor SU(3) limit.

\section*{Acknowledgments}
We thank  members of  PACS Collaboration  for the  gauge configuration
generation.
Lattice  QCD calculations  have been  performed on
the  K computer  at RIKEN,  AICS (Nos.   hp120281, hp130023,  hp140209
hp150223, hp150262, hp160211),
HOKUSAI  FX100 computer  at  RIKEN, Wako  (G15023, G16030)
and HA-PACS at  University of Tsukuba (Nos.14a-20,  15a-30).
We  thank   ILDG/JLDG  \cite{jldg}   which  serves  as   an  essential
infrastructure in this study.
This work  is supported  in part by  MEXT Grand-in-Aid  for Scientific
Research JP25400244,
SPIRE (Strategic Program for Innovative  REsearch) Field 5 project and
``Priority Issue on Post-K  computer'' (Elucidation of the Fundamental
Laws and Evolution of the Universe).


\begin{thebibliography}{99}
\bibitem{Ishii:2006ec}
  N.~Ishii, S.~Aoki and T.~Hatsuda,
  Phys.\ Rev.\ Lett.\  {\bf 99} (2007) 022001
  [nucl-th/0611096].
\bibitem{Aoki:2009ji}
  S.~Aoki, T.~Hatsuda and N.~Ishii,
  Prog.\ Theor.\ Phys.\  {\bf 123} (2010) 89
  [arXiv:0909.5585 [hep-lat]].
\bibitem{Aoki:2011gt}
  S.~Aoki {\it et al.} [HAL QCD Collaboration],
  Proc.\ Japan Acad.\ B {\bf 87} (2011) 509
  [arXiv:1106.2281 [hep-lat]].
\bibitem{Aoki:2012tk}
  S.~Aoki {\it et al.} [HAL QCD Collaboration],
  PTEP {\bf 2012} (2012) 01A105
  [arXiv:1206.5088 [hep-lat]].
\bibitem{Ishikawa:2015rho}
  K.-I.~Ishikawa {\it et al.} [PACS Collaboration],
  PoS LATTICE {\bf 2015} (2016) 075
  [arXiv:1511.09222 [hep-lat]].
\bibitem{HALQCD:2012aa}
  N.~Ishii {\it et al.} [HAL QCD Collaboration],
  Phys.\ Lett.\ B {\bf 712} (2012) 437
  [arXiv:1203.3642 [hep-lat]].
\bibitem{Ishii:2016zsf}
  N.~Ishii {\it et al.},
  PoS LATTICE {\bf 2015} (2016) 087.
\bibitem{Inoue:2011ai}
  T.~Inoue {\it et al.} [HAL QCD Collaboration],
  Nucl.\ Phys.\ A {\bf 881} (2012) 28
  [arXiv:1112.5926 [hep-lat]].
\bibitem{jldg}
http://www.lqcd.org/ildg/ \hspace{2em} and \hspace{2em}
http://www.jldg.org/
\end{thebibliography}
\end{document}